\documentclass{Rinton-P9x6}

\begin{document}

\title{ How much state assignments can differ}

\author{Todd A. Brun}

\address{Institute for Advanced Study, Einstein Drive, Princeton, NJ 08540 \\
Phone:  609-734-8335, FAX:  609-951-4489, email:  tbrun@ias.edu}


\maketitle

\abstracts{
The state that an observer attributes to a quantum system depends on
the information available to that observer.  If two or more observers
have different information about a single system, they will
in general assign different states.  Is there any restriction on what
states can be assigned, given reasonable assumptions about how the
observers use their information?  We derive necessary and sufficient
conditions for a group of general density matrices to characterize
what different people may know about one and the same physical system.  These
conditions are summarized by a single criterion, which we term
{\it compatibility}.}

\section{Observers with differing information}

Suppose there is a physical system ${\cal S}$ with an associated Hilbert
space ${\cal H}_S$ of (finite) dimension $D$, and two observers
Alice and Bob each acquire
information about ${\cal S}$ by some means.  Based on this information,
they describe ${\cal S}$ by two {\it states}:
$\rho_A$ and $\rho_B$, respectively.  If the information that Alice
and Bob acquire is not the same, then in general $\rho_A \ne \rho_B$.

The question we address in this paper is this:  is there any
restriction on the possible assignments $\rho_A$ and $\rho_B$
that can be made?  We argue that there is; and we term two states $\rho_A$
and $\rho_B$ which could represent two descriptions of the same
physical system {\it compatible}\cite{BFM}.

For the purposes of this paper, we assume that all of the information
acquired by Alice and Bob is both {\it accurate} and {\it reliable}.
By {\it accurate} we mean the usual:  that any measurements were performed
and recorded correctly, and that no one deliberately lied to either
of them.  By {\it reliable} we mean something rather more subtle:  that
the information acquired by Alice and Bob has not been rendered
incorrect by disturbances to ${\cal S}$ unknown to them.

Classically this is a rather straightforward assumption, but quantum
mechanically it is not, largely due to the disturbing effects of
measurement.  If Bob, for instance, were to measure ${\cal S}$,
that measurement would disturb the system, potentially introducing
errors into Alice's state assignment. By assuming that the information
of both observers is {\it reliable}, we are explicitly ruling out such
disturbances.  What this means technically will become clear below.

\section{Compatibility}

Intuitively, what should we expect from a criterion for compatibility?
First, if $\rho_A$ and $\rho_B$ are both {\it pure states}, then they
should only be compatible if they are identical:
\begin{equation}
\rho_A = \rho_B = |\psi\rangle\langle\psi|
\label{pure_state}
\end{equation}
for some $|\psi\rangle$.  This seems natural, because pure states
represent states of {\it maximal knowledge} (or minimal ignorance).

Rudolph Peierls\cite{Peierls} suggested two criteria for compatibility
in the general case:
\begin{eqnarray}
{\rm PI:} && [\rho_A,\rho_B] = 0 \;, \nonumber\\
{\rm PII:} && \rho_A \rho_B \ne 0 \;. \nonumber
\end{eqnarray}
These criteria together rule out differing pure state assignments.
Criterion PII seems very natural:  it is just the statement that
$\rho_A$ and $\rho_B$ are not orthogonal, and hence not contradictory.
PI is somewhat less obvious.  It seems to be reasoned by analogy with
compatible observables, in which $[{\hat A},{\hat B}] = 0$ implies
that the observables A and B can be measured simultaneously.

Unfortunately, in the current case it is much too restrictive.
Consider the following example:
\begin{eqnarray}
\rho_A &=& |\psi\rangle\langle\psi| \nonumber\\
\rho_B &=& p |\psi\rangle\langle\psi|
 + (1-p) |\phi\rangle\langle\phi| \;,
\end{eqnarray}
where $\langle\psi|\phi\rangle \ne 0$.
This pair of assignments fails to satisfy PI; but it could easily
arise from two observers with different information.  For instance,
Bob may believe that ${\cal S}$ could have been prepared in either
state $|\psi\rangle$ or $|\phi\rangle$, while Alice has additional
information which rules out the latter.

Let's consider instead one of the following two (equivalent) criteria:
$\rho_A$ and $\rho_B$ are compatible if and only if
there exist decompositions of $\rho_A$ and $\rho_B$
\begin{eqnarray}
\rho_A &=& p_0 |\chi\rangle\langle\chi| +
  \sum_{i>0} p_i |\psi_i\rangle\langle\psi_i| \;,\nonumber\\
\rho_B &=& q_0 |\chi\rangle\langle\chi| +
  \sum_{j>0} q_j |\phi_j\rangle\langle\phi_j| \;,
\label{decomposition}
\end{eqnarray}
which share a state $|\chi\rangle$ in common, such that $p_0,q_0 > 0$;
or equivalently, if and only if
the intersection of their supports is nontrivial,
\begin{equation}
S[\rho_A] \bigcap S[\rho_B] \ne 0 \;,
\label{intersection}
\end{equation}
where $S[\rho]$ is the space spanned by the eigenvectors of $\rho$
with nonzero eigenvalues.
Note that this definition of compatibility extends straightforwardly
to any number of observers, Alice, Bob, Cara, etc.  This criterion
implies PII, and also implies that pure state assignments must be
identical.  We show below that this criterion is both necessary
and sufficient.

\section{Necessity}

Assuming Alice and Bob are rational, if they pool their information
they should agree on a joint state description $\rho_J$.  Furthermore,
since it was assumed that their information was reliable, any measurement
outcome to which either of them initially assigned zero probability must
{\it still} have zero probability in the new state $\rho_J$.

This means that the {\it null space} of the new state $N[\rho_J]$ must
{\it include} the null spaces of the two states $\rho_A$ and $\rho_B$:
\begin{eqnarray}
N[\rho_A] &\subseteq& N[\rho_J] \;, \nonumber\\
N[\rho_B] &\subseteq& N[\rho_J] \;,
\end{eqnarray}
(where $N[\rho]$ is the space spanned by the eigenvectors of $\rho$
with vanishing eigenvalues).  This implies that $N[\rho_J]$ contains
the span of $N[\rho_A]$ and $N[\rho_B]$, which implies in turn that
\begin{equation}
S[\rho_J] \subseteq S[\rho_A] \bigcap S[\rho_B] \;.
\end{equation}

In order for such a joint state assignment to exist, therefore,
the intersection of the supports of $\rho_A$ and $\rho_B$ must be
nontrivial.

\section{Sufficiency}

Obviously, if one is just given two state assignments $\rho_{A,B}$ it
is impossible to know if they are intended to apply to the same system
or not.  So by sufficiency what we mean is that if the assignments
satisfy the compatibility criterion, then they {\it could} be different
descriptions of the same system based on different information.

We prove this by construction.  Suppose $\rho_A$ and $\rho_B$ have
decompositions (\ref{decomposition}).
These state assignments could have arisen in
the following way.  Suppose that there are two ancillary systems
${\cal A}$ and ${\cal B}$ in addition to ${\cal S}$, and that both
Alice and Bob know that the initial state was
\begin{eqnarray}
|\Psi\rangle &=& \frac{1}{N} \Biggl(
  |0\rangle_{\cal A} |0\rangle_{\cal B} |\chi\rangle_{\cal S} +
  \sum_{i>0} \sqrt{\frac{p_i}{p_0}} |0\rangle_{\cal A} |i\rangle_{\cal B}
  |\psi_i\rangle_{\cal S} \nonumber\\
&& + \sum_{j>0} \sqrt{\frac{q_j}{q_0}} |j\rangle_{\cal A} |0\rangle_{\cal B}
  |\phi_j\rangle_{\cal S} \Biggr) \;,
\end{eqnarray}
where $N$ is a normalization factor, $|0\rangle_{\cal A}$ and
$|j\rangle_{\cal A}$ are all mutually orthogonal, and similarly
for $|0\rangle_{\cal B}$ and $|i\rangle_{\cal B}$.
Alice then measures subsystem ${\cal A}$ and Bob measures ${\cal B}$,
both getting result 0, but they do not share their results.  It is
clear that in this case Alice and Bob will make the state assignments
$\rho_A$ and $\rho_B$ given in (\ref{decomposition}).  If they were to
pool their information, they would both arrive at the joint state
assignment $|\chi\rangle$.
So any state assignments which satisfy the compatibility criterion
{\it could} arise from observers with different information about the
same physical system ${\cal S}$.

\section{Further questions}

Both the problem and the solution are easy to state, but lead one
into subtle and interesting questions about
how state assignments are made.  For example:

1.  What restrictions are there on how Alice and Bob acquire their
information, if we want it to be {\it accurate} and {\it reliable}?
How does the case where they perform measurements differ from the case
where they get information from a knowledgeable third party?

2.  If Alice and Bob pool their information, how do they form a joint
state assignment\cite{Jacobs}?  Rather than giving an all or nothing
criterion, is it possible to quantify a {\it degree} of compatibility
between two state assignments\cite{degree}?

3.  Are there other reasonable notions of compatibility?  And
if so, do they lead to well-defined compatibility criteria\cite{CFS}?

All of these questions have been examined to some extent, but much
remains open.  This just shows how far there is to go in completely
understanding state assignment in quantum mechanics.

\section*{Acknowledgments}

First and foremost, I would like to acknowledge my collaborators
Jerry Finkelstein and David Mermin; all technical results of this
paper were developed jointly with them.  I thank Robin Blume-Kohout,
Carl Caves, Jennifer Dodd, Chris Fuchs, David Poulin and R\"udiger
Schack for stimulating conversations and feedback, and the organizers
of QCMC 2002.  This research was supported in part
by the Martin A. and Helen Chooljian Membership in Natural Sciences,
and by DOE Grant No.~DE-FG02-90ER40542.

\end{document}